\documentclass[aps,superscriptaddress,twocolumn,floatfix]{revtex4-1}
\usepackage[utf8]{inputenc}
\usepackage[T1]{fontenc}
\usepackage{mathptmx}
\usepackage{rotating}
\usepackage{longtable}
\usepackage{listings}
\usepackage{xcolor}
\usepackage{siunitx}
\usepackage{lipsum}
\usepackage{wrapfig}
\usepackage{lineno}
\usepackage{pgfplots}
\pgfplotsset{compat=1.17}
\usepgfplotslibrary{external}
\pgfplotsset{plot coordinates/math parser=false}
\usetikzlibrary{plotmarks}
\usepgfplotslibrary{colormaps}
\definecolor{codegreen}{rgb}{0,0.6,0}
\definecolor{codegray}{rgb}{0.5,0.5,0.5}
\definecolor{codepurple}{rgb}{0.58,0,0.82}
\definecolor{backcolour}{rgb}{0.95,0.95,0.92}
\lstdefinestyle{mystyle}{
    backgroundcolor=\color{backcolour},   
    commentstyle=\color{codegreen},
    keywordstyle=\color{magenta},
    numberstyle=\tiny\color{codegray},
    stringstyle=\color{codepurple},
    basicstyle=\ttfamily\footnotesize,
    breakatwhitespace=false,         
    breaklines=true,                 
    captionpos=b,                    
    keepspaces=true,                 
    numbersep=5pt,                  
    showspaces=false,                
    showstringspaces=false,
    showtabs=false,                  
    tabsize=2
}

\usepackage{times,amsmath}
\usepackage{epsfig}
\usepackage{color}
\usepackage{longtable}
\usepackage{ulem}
\usepackage{graphicx}
\usepackage{dcolumn}
\usepackage{bm}
\usepackage{bookmark}
\usepackage{tabularx}
\usepackage{hyperref}
\usepackage{multirow}
\usepackage{booktabs}
\usepackage{tocloft}
\usepackage{pgfplots}
\usepackage{pgfplotstable}
\usepackage{tikz}
\hypersetup{colorlinks=true, citecolor=blue, filecolor=blue, linkcolor=blue, urlcolor=blue}

\usepackage{etoolbox}
\AtBeginEnvironment{table}{\nolinenumbers}
\AtBeginEnvironment{table*}{\nolinenumbers}
\AtBeginEnvironment{figure}{\nolinenumbers}
\AtBeginEnvironment{figure*}{\nolinenumbers}

\begin{document}

\title{Crystal Representation in the Reciprocal Space}

\author{Osman Goni Ridwan}
\affiliation{Department of Mechanical Engineering and Engineering Science, University of North Carolina at Charlotte, Charlotte, NC 28223, USA}

\author{Hongfei Xue}
\affiliation{Department of Computer Science, University of North Carolina at Charlotte, Charlotte, NC 28223, USA}

\author{Youxing Chen}
\affiliation{Department of Mechanical Engineering and Engineering Science, University of North Carolina at Charlotte, Charlotte, NC 28223, USA}

\author{Harish Cherukuri}
\affiliation{Department of Mechanical Engineering and Engineering Science, University of North Carolina at Charlotte, Charlotte, NC 28223, USA}

\author{Qiang Zhu}
\affiliation{Department of Mechanical Engineering and Engineering Science, University of North Carolina at Charlotte, Charlotte, NC 28223, USA}
\affiliation{North Carolina Battery Complexity, Autonomous Vehicle and Electrification (BATT CAVE) Research Center,
Charlotte, NC 28223, USA}

\date{\today}
\begin{abstract}
In crystallography, a structure is typically represented by the arrangement of atoms in the direct space. Furthermore, space group symmetry and Wyckoff site notations are applied to characterize crystal structures with only a few variables. While this representation is effective for data records and human learning, it lacks one-to-one correspondence between the crystal structure and its representation. This is problematic for many applications, such as crystal structure determination, comparison, and more recently, generative model learning. To address this issue, we propose to represent crystals in a four-dimensional (4D) reciprocal space featured by their Cartesian coordinates and scattering factors, which can naturally handle translation invariance and space group symmetry with the help of structure factors. In order to achieve rotational invariance, the 4D coordinates are then transformed into a power spectrum representation under the orthogonal spherical harmonic and radial basis. Hence, this representation captures both periodicity and symmetry of the crystal structure while also providing a continuous representation of the atomic positions and cell parameters in the direct space. Its effectiveness is demonstrated by applying it to several crystal structure matching and reconstruction tasks.
\end{abstract}

\maketitle
\makeatletter

\section{INTRODUCTION}
Crystal structures are commonly represented using direct space coordinates to describe the arrangement of atoms in a periodic lattice. For example, the diamond structure can be expressed as follows.

\begin{table}[h]
\caption{Diamond structure in direct space representation.}
\label{tab:diamond-direct}
\begin{tabular}{cccccc}
\hline\hline
\multicolumn{6}{c}{\textbf{Cell parameters}} \\
 ~~~~$a$ (\AA)~~~~ & ~~~~$b$ (\AA)~~~~ & ~~~~$c$ (\AA)~~~~ & ~~~~$\alpha$ ($^\circ$)~~~~ & ~~~~$\beta$ ($^\circ$)~~~~ & ~~~~$\gamma$ ($^\circ$)~~~~ \\
 3.567 & 3.567 & 3.567 & 90 & 90 & 90\\\hline
\multicolumn{6}{c}{\textbf{Fractional Atomic coordinates}} \\
& Element & $x$ & $y$ & $z$ & \\
& C & 1/8 & 1/8 & 1/8 &  \\
& C & 7/8 & 3/8 & 3/8 &  \\
& C & 1/8 & 5/8 & 5/8 &  \\
& C & 7/8 & 7/8 & 7/8 &  \\
& C & 5/8 & 1/8 & 5/8 &  \\
& C & 3/8 & 3/8 & 7/8 &  \\
& C & 5/8 & 5/8 & 1/8 &  \\
& C & 3/8 & 7/8 & 3/8 &  \\
\hline\hline
\end{tabular}
\end{table}

The cell parameters can also be represented by three lattice vectors $\mathbf{a}_1$, $\mathbf{a}_2$, and $\mathbf{a}_3$ that define the unit cell:

\begin{equation}
\mathbf{L} = [\mathbf{a}_1, \mathbf{a}_2, \mathbf{a}_3]
\end{equation}

This representation provides a straightforward way to describe the atomic arrangement in a crystal lattice.

Another more compact representation is the Crystallographic Information File (CIF) format \cite{CIF}, which further incorporates space group symmetry and Wyckoff site notations to reduce the numerical representation to only a few variables. Using the CIF format, the diamond structure can be represented as space group \textit{Fd}$\bar{3}$\textit{m} (international number: 227), with Wyckoff site 8\textit{a}, where the Wyckoff site is a shorthand notation indicating the multiplicity and the symmetry type (e.g., 8\textit{a}, 4\textit{b}, etc.). Hence, the entire diamond structure can be digitally described by only four elemental components (227, 3.567, 8\textit{a}, C), which is much more compact and easier to process for human readers.

While both representations are convenient for data records and pedagogical human learning, they are not optimal due to the lack of one-to-one correspondence between the structure and its representation. For example, in computer simulations, one may generate the same crystal with a different set of cell parameters and atomic positions, but the resulting structure may still be the same. In fact, using different choices (e.g., origin, cell size, space group symmetry), a diamond can be converted into an infinite list of representations with different space group symmetries. This non-unique mapping may lead to severe confusion and misinterpretation in data analysis.

\begin{table}[h]
\begin{center}
\caption{Diamond structure expressed in different symmetries.}
\label{tab:diamond-sym}
\small  
\begin{tabular}{clll}
\hline\hline
\textbf{Group} & \textbf{Unit Cell} & \textbf{Wyckoff Site 1} & \textbf{Wyckoff Site 2} \\
\midrule
227 & 3.567 & (8a, 1/8, 1/8, 1/8) &  \\
216 & 3.567 & (4a, 0, 0, 0) & (4d, 3/4, 3/4, 3/4) \\ 
210 & 3.567 & (8b, 1/2, 1/2, 1/2) &  \\ 
203 & 3.567 & (8a, 1/8, 1/8, 1/8) &  \\ 
166 & 2.522, 6.178 & (6c, 0, 0, 1/8) & \\
141 & 2.522, 3.567 & (4a, 0, 3/4, 1/8) & \\
122 & 2.522, 3.567 & (4b, 0, 0, 1/2) & \\
119 & 2.522, 3.567 & (2b, 0, 0, 1/2) & (2d, 0, 1/2, 3/4)\\
109 & 2.522, 3.567 & (4a, 0, 0, 7/8) & \\
98 & 2.522, 3.567 & (4b, 0, 0, 1/2) & \\
88 & 2.522, 3.567 & (4a, 0, 1/4, 1/8) & \\
70 & 3.567, 3.567, 3.567 & (8a, 1/8, 1/8, 1/8) & \\
$\cdots$ & $\cdots$ & $\cdots$ & $\cdots$ \\
\hline\hline
\end{tabular}
\end{center}
\end{table}

Furthermore, this ambiguity can be particularly problematic for new applications such as generative model based materials design. In these tasks, one aims to find a smooth latent representation where similar crystal structures can be locally clustered and represented by a unique set of variables regardless of choices on space group symmetry, as well as the translational and rotational invariance. However, due to the non-unique representation in direct space, the generative model may struggle to learn a truly smooth latent space, leading to overfitting and poor generalization. As a result, they attempt to generate structures that are either nearly identical to the training data or simply nonphysical at all.

In crystallography, a more robust representation is the 4D reciprocal space featured by its Cartesian coordinates and corresponding scattering factors. This representation takes advantage of the structure factor concept to naturally handle translation invariance and space group symmetry. However, the 4D reciprocal space representation suffers from the rotational invariance issue, meaning that the same structure can be represented by different variables depending on its orientation in space. 
To address this challenge, we propose to transform the 4D coordinates into a power spectrum using orthogonal spherical harmonics and radial basis functions. Hence, the power spectrum representation is not only translation and rotation invariant, but also encodes structural angular information in different channels, thus it is useful for tasks such as crystal structure matching and generative model learning.

The remainder of this paper is organized as follows: Section~\ref{sec2} provides background on electron scattering and structure factors, establishing the foundation for our approach. Section~\ref{sec3} introduces how power spectrum are used to characterize crystal structures in 4D reciprocal space. Section~\ref{sec4} demonstrates the practical effectiveness of this representation through numerical experiments on structural similarity comparison and crystal structure reconstruction tasks, followed by a conclusion in Section~\ref{sec5}. 

\section{Background}\label{sec2}
\subsection{Electron Scattering of an Isolated Atom}
In an isolated atom, the surrounding electrons can be treated as a continuous charge distribution $\rho(\mathbf{r})$, which gives the probability of finding an electron at position $\mathbf{r}$ in space. One can inversely probe the distribution by using the incident electron wave. The scattering process is governed by the interaction between the incident wave and the electron density distribution. 
Using a list of orthogonal plane waves to probe the $\rho(\mathbf{r})$, this is essentially a Fourier transform of the electron density distribution $\rho(\mathbf{r})$:

\begin{equation}\label{eq:af}
f(\mathbf{q}) = \int \rho(\mathbf{r}) e^{i\mathbf{q}\cdot\mathbf{r}} d^3r
\end{equation}

In a single electron with a spherically spread distribution, the scattering amplitude is maximum at the forward direction ($|\mathbf{q}|$ = 0) and decreases at shorter wavelengths (i.e., when $|\mathbf{q}|$ increases). For an atom containing $Z$ electrons, the scattering amplitude equals $Z$ times that of a single electron when it is at the forward direction. However, this proportionality does not hold for other scattering directions due to interference effects between electrons. To quantify the scattering efficiency of an atom in any direction, the real-valued atomic scattering factor $f(\mathbf{q})$ is defined as:

\begin{equation}
f = \frac{\text{Amplitude of the wave scattered by the atom}}
{\text{Amplitude of the wave scattered by a free electron}}
\end{equation}

More detailed explanation can be found in Ref.~\cite{Cullitybook}. In practice, $f(\mathbf{q})$ is complicated to derive analytically, but can be tabulated for different elements based on a sum of four Gaussians with different coefficients $a_i$, $b_i$, and $c$ as follows \cite{brown2006intensity}: 

\begin{equation}\label{eq-table}
f(\mathbf{q}) = \sum_{j=1}^{4} a_i \exp 
\left( -b_i \left( \frac{|\mathbf{q}|^2}{4} \right) \right) + c
\end{equation}

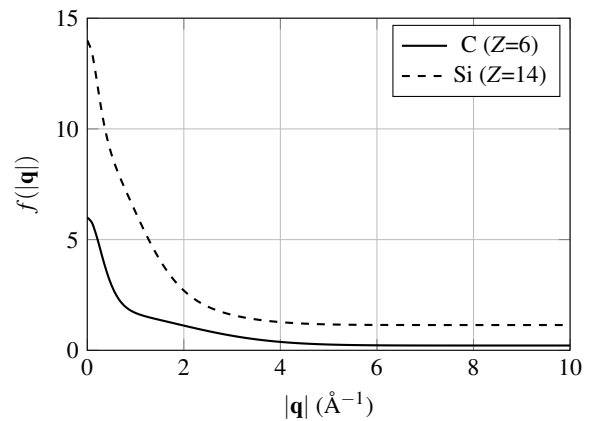
\begin{figure}[h]
\centering
\begin{tikzpicture}
\begin{axis}[
    xlabel={$|\mathbf{q}|$ (\AA$^{-1}$)},
    ylabel={$f(|\mathbf{q}|)$},
    xmin=0, xmax=10,
    ymin=0, ymax=15,
    grid=major,
    width=8cm,
    height=6cm
]
\addplot[thick,smooth,domain=0:10,samples=100] {
    2.31*exp(-20.8439*x^2/4) + 
    1.02*exp(-10.2075*x^2/4) + 
    1.5886*exp(-0.5687*x^2/4) + 
    0.865*exp(-51.6512*x^2/4) + 0.2156
};
\addlegendentry{C ($Z$=6)}

\addplot[thick,smooth,dashed,domain=0:10,samples=100] {
    6.2915*exp(-2.4386*x^2/4) + 
    3.0353*exp(-32.3337*x^2/4) + 
    1.9891*exp(-0.6785*x^2/4) + 
    1.541*exp(-81.6937*x^2/4) + 1.1407
};
\addlegendentry{Si ($Z$=14)}
\end{axis}
\end{tikzpicture}
\caption{Atomic form factors $f(|\mathbf{q}|)$ as functions of scattering vector magnitude for C and Si atoms according to Ref. \cite{brown2006intensity}.}
\label{fig:af}
\vspace{-3mm}
\end{figure}

Fig. \ref{fig:af} displays two examples for neutral C and Si atoms, respectively, by using the tabulated data. In short, $f$ should be equal to $Z$ for any atom scattering when $\mathbf{q} = 0$. As $|\mathbf{q}|$ increases, the scattering amplitude $f(\mathbf{q})$ decreases due to the finite size of the atom and the distribution of its electrons. This decay is typically exponential, reflecting the fact that the electron density falls off with distance from the nucleus. 

\subsection{Electron Scattering of a Crystal}

Now, we consider many atoms forming a regular crystal in 3D space. In this case, we need to consider coherent scattering, not from an isolated atom, but from all atoms in the entire crystal. Hence, we add the scattered wave contributions from all atoms in the unit cell for each wave vector $\mathbf{q}$. 

In different directions, the scattering on each atom will interfere with each other, leading to a complex scattering pattern. Hence, we define the sum of the contributions $F$ from all atoms in the unit cell, weighted by their atomic form factors and phase factors:

\begin{equation}\label{eq:sf1}
F(\mathbf{q}) = \sum_{j=1}^N f_j(\mathbf{q}) e^{i\mathbf{q}\cdot\mathbf{r}_j}
\end{equation}

where the sum runs over all atoms $j$ in the unit cell, $f_j(\mathbf{q})$ is the atomic form factor of atom $j$ as defined in Eq. \ref{eq-table}. 

Unlike the isolated atom case, the choice of $\mathbf{q}$ is not arbitrary, but is determined by the crystal lattice periodicity. Due to the lattice constraint, the $\mathbf{q}$ vector must be a linear combination of the reciprocal lattice vectors $\mathbf{G}$ as follows:

\begin{equation}
\mathbf{q}_{hkl} = h \mathbf{b}_1 + k \mathbf{b}_2 + l \mathbf{b}_3
\end{equation}

where $h, k, l$ are integers (Miller indices) and $\mathbf{b}_1, \mathbf{b}_2, \mathbf{b}_3$ are the reciprocal lattice vectors corresponding to the direct lattice vectors $\mathbf{a}_1, \mathbf{a}_2, \mathbf{a}_3$. 

Thus, Eq. \ref{eq:sf1} is also commonly written in terms of the Miller indices $(h, k, l)$ and fractional atomic coordinates as follows:
\begin{equation}\label{eq:sf2}
F_{hkl} = \sum_{j=1}^N f_j \exp[-2\pi i (h x_j + k y_j + l z_j)]
\end{equation}

where $(x_j, y_j, z_j)$ are the fractional coordinates of atom $j$ in the unit cell.

The quantity $F_{hkl}$ is known as the structure factor, which is is a complex number to describe the amplitude and phase of the scattered wave in the direction of the reciprocal lattice vector $\mathbf{q}_{hkl}$. For practical applications, it is easy to measure the magnitude but hard to extract the phase information. Therefore, we are mostly interested in the intensity of the scattered wave, which is given by the square of the magnitude of the structure factor:

\begin{equation}
I(\mathbf{q}) = |F(\mathbf{q})|^2 
\end{equation}

At $\mathbf{q}$ = 0, the intensity $I(\mathbf{q})$ is equal to the total number of electrons, which is proportional to the number of atoms in the unit cell. As $|\mathbf{q}|$ increases, the intensity $I(\mathbf{q})$ typically decreases due to the decay behavior of the atomic form factor as discussed earlier. For the convenience of calculation, we can further define the normalized intensity $\bar{I}(\mathbf{q})$ as 

\begin{equation}\label{eq:sf3}
\bar{I}(\mathbf{q}) = \frac{I(\mathbf{q})}{I(\mathbf{q}=0)} = \frac{|F(\mathbf{q})|^2}{|F(\mathbf{q}=0)|^2}
\end{equation}

In X-ray diffraction experiments, a polarization factor needs to be applied to correct the intensity measurements \cite{Cullitybook}. However, since this correction is not relevant to the current discussion of crystal representation, we will omit this factor in the following sections.

\subsection{4D Reciprocal Space Representation and Its Properties}

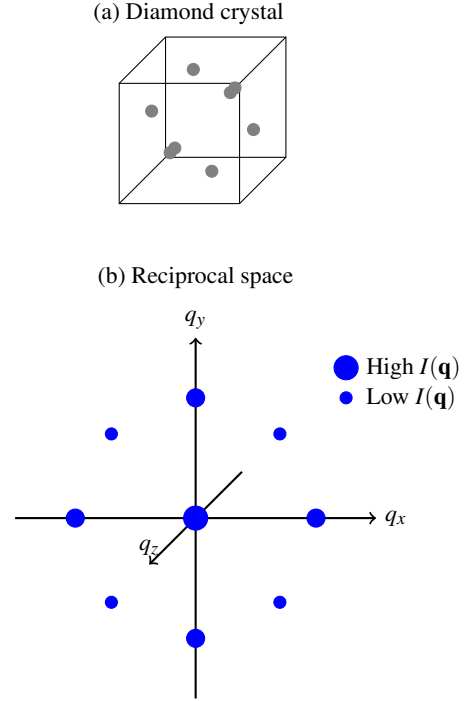
\begin{figure}[h]
\centering
\begin{tikzpicture}[scale=0.8]
        \begin{scope}[yshift=4cm, xshift=-0.5cm]
            \begin{scope}[yshift=4cm, rotate=0] 
                \draw[black] (0,0,0) -- (2,0,0) -- (2,2,0) -- (0,2,0) -- cycle;
                \draw[black] (0,0,2) -- (2,0,2) -- (2,2,2) -- (0,2,2) -- cycle;
                \draw[black] (0,0,0) -- (0,0,2);
                \draw[black] (2,0,0) -- (2,0,2);
                \draw[black] (2,2,0) -- (2,2,2);
                \draw[black] (0,2,0) -- (0,2,2);
                
                \filldraw[gray] (1/4, 1/4, 1/4) circle (0.1);
                \filldraw[gray] (7/4, 3/4, 3/4) circle (0.1);
                \filldraw[gray] (1/4, 5/4, 5/4) circle (0.1);
                \filldraw[gray] (7/4, 7/4, 7/4) circle (0.1);
                \filldraw[gray] (5/4, 1/4, 5/4) circle (0.1);
                \filldraw[gray] (3/4, 3/4, 7/4) circle (0.1);
                \filldraw[gray] (5/4, 5/4, 1/4) circle (0.1);
                \filldraw[gray] (3/4, 7/4, 3/4) circle (0.1);
                
                \node at (0, 2, -1) {(a) Diamond crystal};
            \end{scope}
        \end{scope}

        \begin{scope}[yshift=2cm]
            \draw[->,thick] (-3,0,0) -- (3,0,0) node[right] {$q_x$};
            \draw[->,thick] (0,-3,0) -- (0,3,0) node[above] {$q_y$};
            \draw[->,thick] (0,0,-2) -- (0,0,2) node[above] {$q_z$};
            \filldraw[blue] (0,0,0) circle (0.2);
            \filldraw[blue] (2,0,0) circle (0.15);
            \filldraw[blue] (-2,0,0) circle (0.15);
            \filldraw[blue] (0,2,0) circle (0.15);
            \filldraw[blue] (0,-2,0) circle (0.15);
            \filldraw[blue] (1.4,1.4,0) circle (0.1);
            \filldraw[blue] (-1.4,1.4,0) circle (0.1);
            \filldraw[blue] (1.4,-1.4,0) circle (0.1);
            \filldraw[blue] (-1.4,-1.4,0) circle (0.1);
            
            \filldraw[blue] (2.5,2.5) circle (0.2) node[right,black] {~~High $I(\mathbf{q})$};
            \filldraw[blue] (2.5,2) circle (0.1) node[right,black] {~~Low $I(\mathbf{q})$};
            \node at (0, 4) {(b) Reciprocal space};

        \end{scope}
\end{tikzpicture}
\caption{The diamond crystal's representations at the (a) real and (b) reciprocal space. In (b), the size of each point represents the relative intensity $I(\mathbf{q})$ at that reciprocal lattice vector $\mathbf{q}$.}
\vspace{-3mm}
\label{fig:4d-reciprocal}
\end{figure}

Using Eq. \ref{eq:sf3}, one can compute the intensity $\bar{I}$ for a series of $hkl$ indices of a crystal. This information can be mapped to a 4D reciprocal space representation, defined by the Cartesian coordinates of the reciprocal lattice vector $\mathbf{q} = (q_x, q_y, q_z)$ and the corresponding intensity $\bar{I}(\mathbf{q})$. Fig. \ref{fig:4d-reciprocal} illustrates this representation for the diamond structure in Table \ref{tab:diamond-direct}. The physical interpretation is that it describes the scattering pattern of periodically arranged spherical electron clouds in the crystal lattice. In a real crystalline solid, the electron clouds are not perfectly spherical, but rather have a more complex shape due to the interactions between electrons and the atomic nuclei \cite{COHEN197037, Chelikowsky1974electronic}. However, this simplified assumption still retains rich information about the periodicity and symmetry of the crystal structure that can be used for compact characterization. Below are some key properties of the 4D reciprocal space representation.

\vspace{2mm}
\noindent
\textbf{1. Translation Invariance.} The structure factor $F(\mathbf{q})$ captures the crystal lattice periodicity through the reciprocal lattice vectors $\mathbf{q}_{hkl}$. Two types of translational transformations can be applied to the crystal structure: (1) translations within the unit cell, e.g., [$\frac{1}{2}(\mathbf{a}_1+\mathbf{a}_2), \frac{1}{2}(\mathbf{a}_1+\mathbf{a}_2), \mathbf{a}_3]$ and (2) translations involving cell size changes, e.g., $[2\mathbf{a}_1, \mathbf{a}_2, \mathbf{a}_3]$. The former results in the same set of coordinates and intensities, just labeled by different $hkl$ indices. The latter may add or remove $\mathbf{q}$ points when the cell size changes, but these added/removed points have zero intensity ($I(\mathbf{q})$ = 0). After removing zero-intensity points, the 4D reciprocal space representation remains invariant under translational symmetry operations.

\vspace{2mm}
\noindent
\textbf{2. Convergence.} Due to the decay behavior of the atomic form factor $f(\mathbf{q})$ with increasing $|\mathbf{q}|$, the intensity $I(\mathbf{q})$ eventually becomes negligible at large $|\mathbf{q}|$, similar to the kinetic energy cutoff concept in plane wave based electronic structure calculations \cite{martin2020electronic}. This means that structure comparisons can be made using only a finite set of low-index $hkl$ points in 4D space, making this representation particularly suitable for data analysis and machine learning applications.

\vspace{2mm}
\noindent
\textbf{3. Smoothness.} When atomic positions are slightly modified, it leads to changes in the phase factor $e^{i\mathbf{q}\cdot\mathbf{r}}$ in Eq \ref{eq:sf2}, affecting the intensity $I(\mathbf{q})$ without changing the Cartesian coordinates of the $\mathbf{q}_{hkl}$. Conversely, when cell parameters are adjusted (e.g., cell size or angles), it primarily impacts the Cartesian coordinates of the reciprocal lattice vectors $\mathbf{q}_{hkl}$. In both cases, small structural changes should result in proportionally small changes in the $\mathbf{q}_{hkl}$ coordinates and intensity $I(\mathbf{q})$, making this representation moderately sensitive to minor structural variations while maintaining overall stability in the data.

\vspace{3mm}
\noindent
\textbf{4. Reconstruction Capability}. The reciprocal space representation contains sufficient information to reconstruct the original crystal structure, even though phase information is lost when converting structure factors to intensities. The atomic positions and cell parameters can be recovered from intensity data using established crystallographic techniques like Patterson or direct methods \cite{glusker2010crystal}. This demonstrates that the 4D reciprocal space representation completely captures the structural information of crystals.

Different from previously proposed atomic local environment descriptors ~\cite{Oganov-JCP-2009, Behler-PRL-2007, Bartok-PRB-2013, thompson2015spectral,samanta2018representing, MTP, ace2019, zhu2016fingerprint, kondor2018clebsch,  musil2021physics, nigam2022unified} which primarily probes the atomic environment within a short cutoff distance, this 4D 
the reciprocal space representation effectively captures the crystal periodicity and symmetry in a long range, thus providing a more complete description on both atomic positions and cell parameters of a crystal structure. It is translation invariant, convergent, smooth, and capable of reconstructing the original crystal structure from intensity data. These properties make it a suitable choice for crystallographic data analysis, such as structure matching and generative model learning.

\section{Rotation Invariant Representation}\label{sec3}
While these 4D coordinates can effectively capture the crystal periodicity and symmetry, they still suffer from the rotational invariance issue. Ideally, we would like to have a rotation-invariant representation, meaning that the same crystal structure can be represented by the same variables regardless of its orientation in space. Below, we will start with the radial distribution function and then introduce the power spectrum representation to address this issue.

\subsection{Radial Distribution Function}
To capture the rotational invariance, one natural approach is to use the radial distribution function (RDF), which describes how the density of reference points varies as a function of distance from a reference. For the 4D reciprocal space representation, the RDF can be computed by summing the intensity $I(\mathbf{q})$ over all directions at each binned distance $|\mathbf{q}| = d$:

\begin{equation}
g(d) = \sum_{\mathbf{q}: |\mathbf{q}| = d} I(\mathbf{q}) 
\end{equation}

In fact, $g(d)$ is very similar to the commonly used 
in experimental powder X-ray diffraction (PXRD) pattern, except that i) the PXRD has an additional polarization factor correction and ii) the PXRD is normalized to the strongest intensity at the nonzero $\mathbf{q}$ point. 
While it provides a rotation invariant representation, this function may group symmetrically nonequivalent $(hkl)$ planes into a single $d$ bin, causing two different structures to exhibit very similar $g(d)$ spectra. Additionally, small structural perturbations can significantly alter the $g(d)$ spectrum, making it difficult to detect structural similarities. Therefore, it is necessary to preserve more angular information by doing better feature engineering.

\subsection{Spherical Harmonics Expansion}
The challenge of representing 4D reciprocal space is analogous to shape matching in computer vision \cite{kazhdan2003rotation, kazhdan2004shape,Kondor2007} and machine learning force field development \cite{Behler-PRL-2007, Bartok2009, thompson2015spectral, wood2018extending, MTP, ace2019, yanxon2020pyxtalff,kondor2018clebsch,nigam2022unified,musil2021physics}. Inspired by these prior works, we use spherical harmonics $Y_{lm}(\theta, \phi)$, a set of orthogonal functions defined on the surface of a sphere, to represent 3D point clouds in a rotation-invariant manner. In the spherical harmonics, the degree $l$ and order $m$ are given by:

\begin{equation}
Y_{lm}(\theta, \phi) = N_{lm} P_l^m(\cos \theta) e^{im\phi}
\end{equation}

where $N_{lm}$ is a normalization constant, $P_l^m$ are the associated Legendre polynomials, and $(\theta, \phi)$ are the polar and azimuthal angles on the sphere. The indices $l$ and $m$ are non-negative integers, where $l$ is the degree and $m$ is the order of the spherical harmonic. 

For a function $f(\theta, \phi)$ defined on a unit sphere, it can be expanded in terms of spherical harmonics as follows:

\begin{equation}
f(\theta, \phi) = \sum_{l=0}^{\infty} \sum_{m=-l}^{l} a_{lm} Y_{lm}(\theta, \phi)
\end{equation}

where $a_{lm}$ are the coefficients of the expansion, which can be computed as:

\begin{equation}
a_{lm} = \int_0^{\pi} \int_0^{2\pi} f(\theta, \phi) Y_{lm}^*(\theta, \phi) \sin \theta d\theta d\phi,
\end{equation}
where $Y_{lm}^*$ is the complex conjugate of the spherical harmonic function. The coefficients $a_{lm}$ capture the angular dependence of the function $f(\theta, \phi)$ on the sphere.

\subsection{Orthonormal Radial Basis}
Next, the radial dependence of the reciprocal space representation can be captured by a set of radial basis functions $R(d)$ defined on the radial distance $d = |\mathbf{q}|$. In the context of atomistic representations and machine-learning force fields, various functional forms have been proposed for radial basis expansion, including Gaussian functions \cite{Behler-PRL-2007}, B-splines \cite{xie2023ultra}, Bessel functions \cite{Bartok-PRB-2013}, and Chebyshev polynomials \cite{goldman2021semi}. In this work, we primarily employ an orthonormal spherical Bessel basis that is consistent with the three-dimensional spherical harmonic expansion used for the angular dependence. Chebyshev polynomials are considered as an alternative functional basis.

For the spherical Bessel basis, we construct radial functions that are orthonormal under the three-dimensional radial measure $r^2\,dr$:
\begin{equation}\label{eq:orthonormality}
\int_0^{d_{\text{max}}} r^2 R_n(r) R_m(r) \, dr = \delta_{mn}.
\end{equation}

The spherical Bessel radial basis functions are defined as
\begin{equation}\label{eq:basis}
R_n(d) = N_n\, j_0\!\left(z_n \frac{d}{d_{\text{max}}}\right),
\end{equation}
where $j_0$ is the spherical Bessel function of the first kind at order zero, and $z_n$ denotes the $n$-th positive zero of $j_0$. The normalization constant is given by
\begin{equation}
N_n = \frac{\sqrt{2}}{d_{\text{max}}^{3/2} \, |j_1(z_n)|},
\end{equation}
which ensures the orthonormality condition in Eq.~\ref{eq:orthonormality}. 

Chebyshev polynomials are defined as
\begin{equation}
T_n(d) = \cos\!\left[n \, \arccos\!\left(2\frac{d}{d_{\text{max}}} - 1\right)\right],
\end{equation}
and form an orthogonal basis on the interval $[-1,1]$ with respect to the weight $(1-x^2)^{-1/2}$ after the linear mapping $x = 2d/d_{\text{max}} - 1$. 

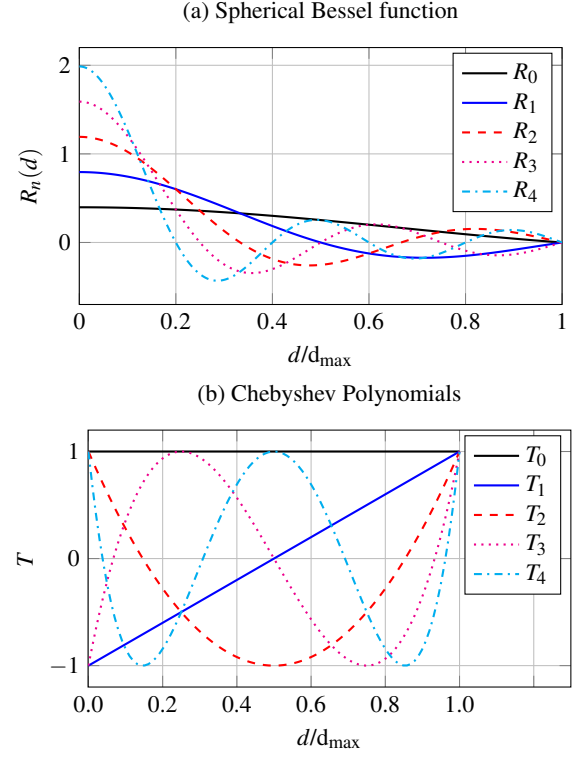
\begin{figure}[h]
\centering
\begin{tikzpicture}
\begin{axis}[
    xlabel={$d$/d$_{\text{max}}$},
    ylabel={$R_n(d)$},
    xmin=0, xmax=1,
    ymin=-0.7, ymax=2.2,
    grid=major,
    width=8cm,
    height=5cm,
    xtick={0, 0.2, 0.4, 0.6, 0.8, 1.0},
    title={(a) Spherical Bessel function}
]

\addplot [black, thick, smooth] table [x index=0, y index=1] {Fig3-bessel.txt}; \addlegendentry{$R_0$}
\addplot [blue, thick, smooth] table [x index=0, y index=2] {Fig3-bessel.txt}; \addlegendentry{$R_1$}
\addplot [red, thick, dashed] table [x index=0, y index=3] {Fig3-bessel.txt}; \addlegendentry{$R_2$}
\addplot [magenta, thick, dotted] table [x index=0, y index=4] {Fig3-bessel.txt}; \addlegendentry{$R_3$}
\addplot [cyan, thick, dash dot] table [x index=0, y index=5] {Fig3-bessel.txt}; \addlegendentry{$R_4$}
\end{axis}
\end{tikzpicture}
\hspace{0.5cm}
\begin{tikzpicture}
\begin{axis}[
    xlabel={$d$/d$_{\text{max}}$},
    ylabel={$T$},
    xmin=-0.01, xmax=13,
    ymin=-1.2, ymax=1.2,
    grid=major,
    width=8cm,
    height=5cm,
    xticklabels={0.0,0.0,0.2,0.4,0.6,0.8,1.0},
    title={(b) Chebyshev Polynomials}
]
\addplot[black,thick,smooth,domain=0:10,samples=100] {1};
\addplot[blue,thick,smooth,domain=0:10,samples=100] {2*x/10-1};
\addplot[red,thick,dashed,domain=0:10,samples=100] {2*(2*x/10-1)^2-1};
\addplot[magenta,thick,dotted,domain=0:10,samples=100] {4*(2*x/10-1)^3-3*(2*x/10-1)};
\addplot[cyan,thick,dash dot,domain=0:10,samples=100] {8*(2*x/10-1)^4-8*(2*x/10-1)^2+1};
\legend{$T_0$, $T_1$,$T_2$,$T_3$,$T_4$}
\end{axis}
\end{tikzpicture}
\vspace{-3mm}
\caption{Comparison of spherical Bessel function (top) and Chebyshev polynomials (bottom) used as radial basis functions.}
\label{fig:basis}
\end{figure}

Figure~\ref{fig:basis} compares representative spherical Bessel radial modes and Chebyshev polynomials. The spherical Bessel functions exhibit increasing radial oscillations with mode index $n$ and satisfy a hard cutoff at $d = d_{\text{max}}$, reflecting their interpretation as standing spherical waves. In contrast, Chebyshev polynomials are bounded, globally supported, and oscillatory over the entire interval. These two bases therefore provide different inductive biases for encoding radial structural information.

Using the radial basis, the expansion coefficients $a_{lm}$ can be further weighted by the radial distance $d$ as follows:

\begin{equation}
A_{nlm}(d) = \int_0^{\pi} \int_0
^{2\pi} f(\theta, \phi) Y_{lm}^*(\theta, \phi) R_n(d) \sin \theta d\theta d\phi
\end{equation}

where $R_n$ is the radial basis function defined in Eq \ref{eq:basis}. The resulting complex-valued coefficients $A_{nlm}(d)$ can be used to fully describe the 4D reciprocal space representation in a rotation-invariant manner. However, these coefficients can still change upon the rotation.

\subsection{Power Spectrum Representation}
To achieve rotational invariance from $A_{nlm}$, we can transform it into a power spectrum representation, following the pioneering work by Steinhardt \cite{steinhardt1983bond} and subsequent developments in machine learning force fields \cite{Bartok2009, Bartok-PRB-2013, thompson2015spectral, wood2018extending, MTP, ace2019,yanxon2020pyxtalff}. The power spectrum is defined as the sum of the squared magnitudes of the coefficients $A_{nlm}(d)$ over all orders $m$ for a given degree $l$:

\begin{equation}\label{eq:ps}
    P_{nl} = \sum^{\text{all ds}}_i\sum_{m=-l}^{l} |A_{nlm}(d_i)|^2
\end{equation}

This power spectrum $P_{nl}$ is a real-valued 2D array that captures the rotationally invariant features of the 4D reciprocal space representation. It can be interpreted as a measure of the intensity distribution of the crystal structure in the spherical harmonics basis, weighted by the radial basis functions. 
We note that this formula may be further extended to include the cross-radial terms $P_{nn`l}$ to express radial–radial correlations, which will be left for future studies.

Using Eq.~\ref{eq:ps}, we finally construct a two-index array $P_{nl}$ to describe the crystal structure regardless its orientation, where $n$ is the index corresponding to the radial basis function to capture the radial feature and $l$ is the degree of the spherical harmonics to detect the angular information. 

\subsection{Practical Power Spectrum Computation}
Hence, the power spectrum can be computed as follows:
\begin{enumerate}
    \item For each crystal structure, compute the Cartesian coordinates of the reciprocal lattice vectors $\mathbf{q}_{hkl}$ for all relevant Miller indices $(h, k, l)$ with a $d_\text{max}$.
    \item Compute the structure factor $F_{hkl}$ for all relevant Miller indices $(h, k, l)$ using eq \ref{eq:sf2}.
    \item Compute the intensity $I_{hkl} = |F_{hkl}|^2$ and normalize it to obtain $\bar{I}_{hkl}$ using eq \ref{eq:sf3}.
    \item Compute the spherical harmonics coefficients $A_{nlm}(d)$ for each $d_{h kl}$ using the radial basis functions $R_n(d)$ and the spherical harmonics $Y_{lm}(\theta, \phi)$.
    \item Filer out the $q_{hkl}$ with zero intensities.
    \item Compute the power spectrum $P_{nl}$ using Eq \ref{eq:ps}.
\end{enumerate}
The optimal choice of $d_\text{max}$, radial basis functions, and spherical harmonics degree can be determined based on the specific application. A typical setting is to use $d_\text{max}$ = 10 \AA$^{-1}$, with the first 10 Bessel functions ($n_\text{max}$=10) and spherical harmonics up to degree 10 ($l_\text{max}$=10) should be sufficient. This setting provides a good balance between computational efficiency and representation accuracy for most crystal structures.

After $P_{nl}$ is obtained, the mismatch metric between two crystal structures can be computed using the L2-distance between their power spectrum representations:

\begin{equation}
D(s_1, s_2) = \sqrt{\sum_{nl} (P^{s_1}_{nl} - P^{s_2}_{nl})^2}
\end{equation}

This metric can be used to quantify similarity or dissimilarity. If two structures are identical, $D(s_1, s_2)$ should be zero. If they are dissimilar, the distance should be large. 

\begin{figure}[ht]
    \centering
\includegraphics[width=0.48\textwidth]{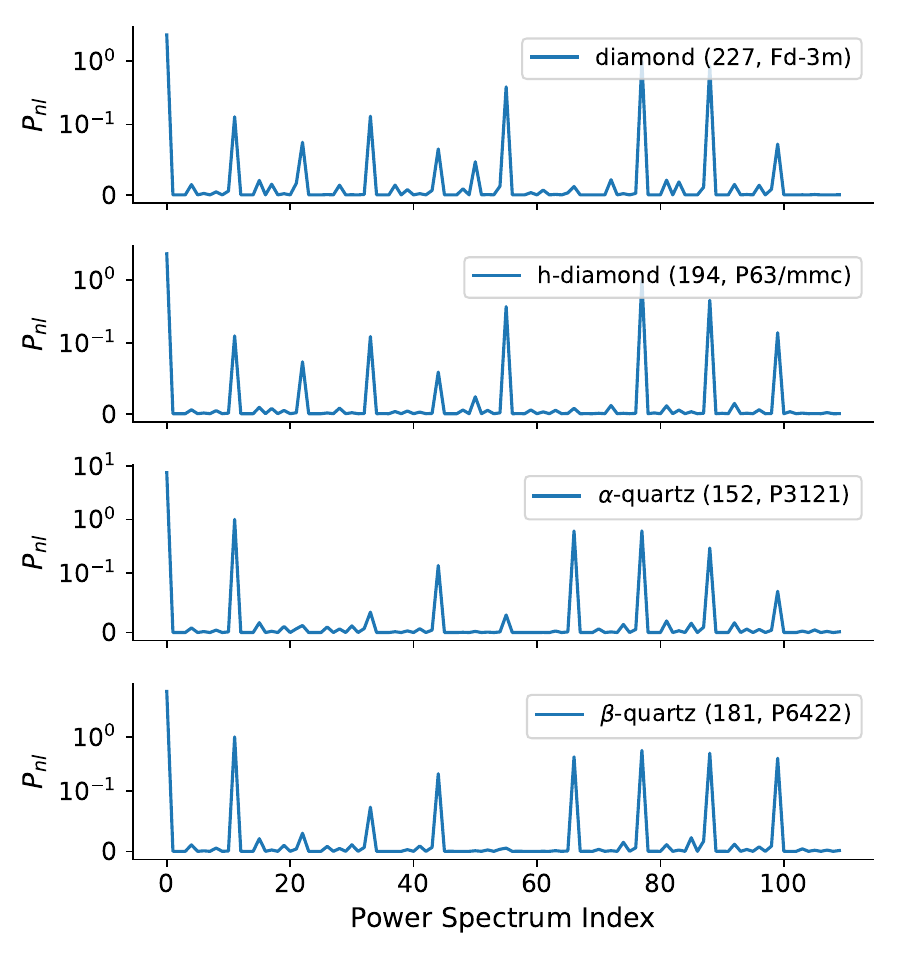}
\vspace{-5mm}
    \caption{The computed power spectrum $P_{nl}$ for various representative crystals. Note that $P_{nl}$ is intrinsically a 1D array indexed by $(n,l)$; the spread of peaks in the line plot is merely a visualization artifact and does not carry physical meaning.}
    \label{fig:p-demo}
    \vspace{-3mm}
\end{figure}

Fig. \ref{fig:p-demo} displays the computed $P_{nl}$ arrays for representative crystals, including cubic/hexagonal diamond, and $\alpha$/$\beta$ quartz, based on $l_\text{max}$ = 10 and $n_\text{max}$ = 10 for using the Bessel functions as the radial basis. The similarity between cubic and hexagonal diamond power spectra, as well as between $\alpha$/$\beta$ quartz spectra, demonstrates that the representation captures underlying structural relationships despite different space group symmetries. To avoid dominance of the leading $P_{00}$ term in visualization, we normalize each spectrum by the maximum magnitude of higher-order components ($n>0$), enhancing contrast among physically informative modes. 

The numerical implementation of the power spectrum representation can be found in the \texttt{PyXtal} package \cite{pyxtal}, which provides a convenient interface for computing the power spectrum representation and comparing crystal structures. 

\begin{figure*}[ht]
    \centering
    \includegraphics[width=0.98\textwidth]{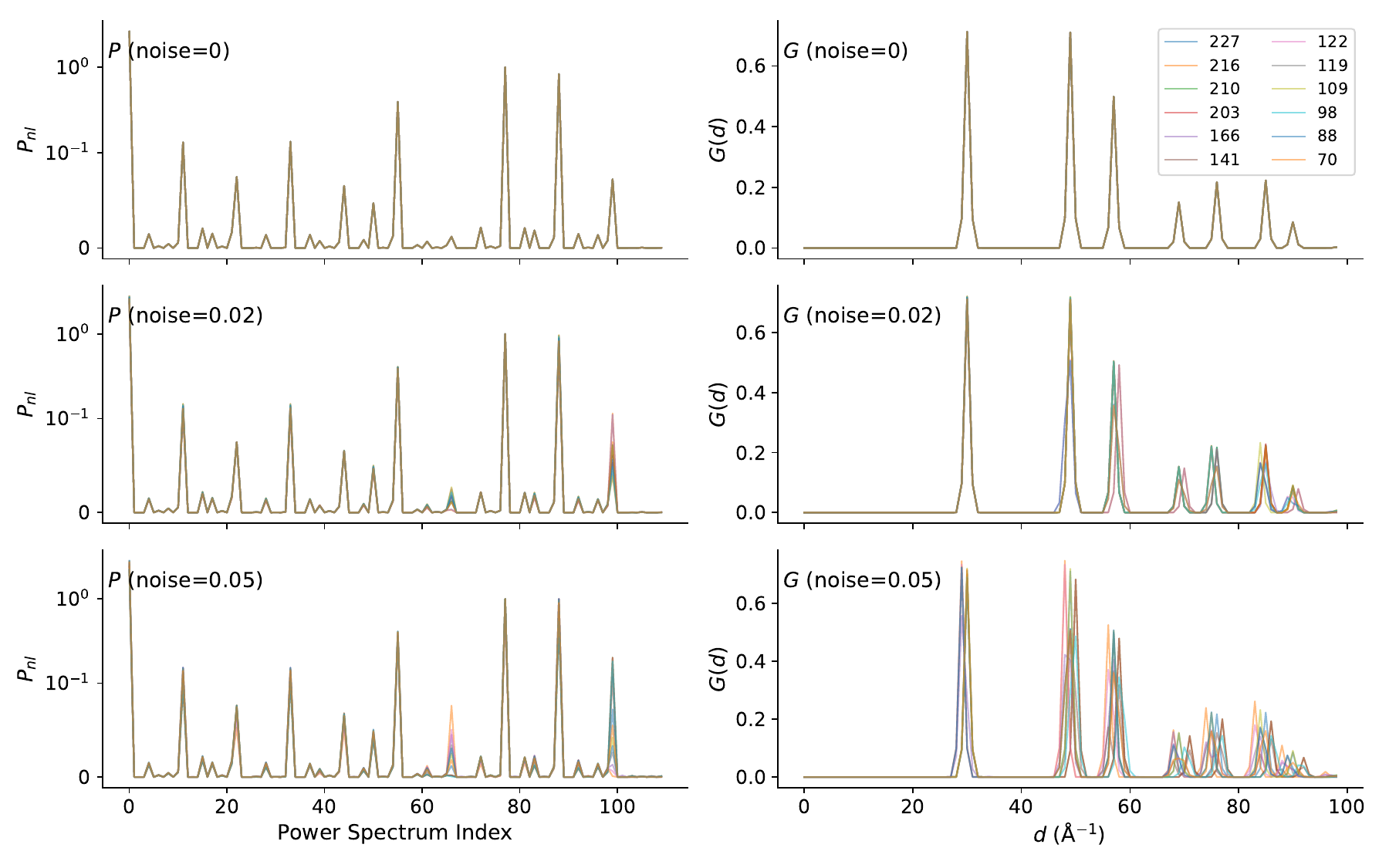}
    \vspace{-4mm}
    \caption{The computed power spectrum $P_{nl}$ (left) and radial distribution function $G(d)$ (right) for various diamond structures listed in Table \ref{tab:diamond-sym}. The $P_{nl}$ is computed using the Bessel functions as the radial basis functions and spherical harmonics up to degree 10. The $G(d)$ is computed by summing the intensity over all directions at each binned distance $d$. The noise levels are indicated in the legend.}
    \label{fig:diamond-ps}
    \vspace{-3mm}
\end{figure*}

\section{Numerical Performances}\label{sec4}
Using the power spectrum representation, we aim to understand: (i) if the representation can robustly capture structural similarity in the presence of noise and errors; and (ii) if crystal structures can be reconstructed given target power spectrum patterns $P_{nl}$. The following subsections present two numerical experiments addressing these questions.

\subsection{Crystal Matching on Diamond Structures}
We consider the diamond structure with different representations as listed in Table \ref{tab:diamond-sym}. To test the sensitivity of the power spectrum representation, we consider adding random noise to the atomic positions and cell parameters of the diamond structure with two levels (0.02~\AA~and 0.05~\AA). The noise is uniformly distributed within a certain range, which can be adjusted to control the level of perturbation. Then, we plot the $P_{nl}$ and $G(d)$ representations of these structures in Fig. \ref{fig:diamond-ps}.

Without adding the noise, the top panels of Fig.~\ref{fig:diamond-ps} show that both $P_{nl}$ and $G(d)$ are identical for the whole 12 diamond structures, despite the fact that the structures are expressed by different space group symmetries. In both plots, they are featured by a list of distinct peaks corresponding to different power spectrum indices or $d$ values.

When noises are added to these structures (see the middle and bottom panels of Fig.~\ref{fig:diamond-ps}), the $P_{nl}$ representation remains more stable and retains the main structural features even at a high noise level of 0.05~\AA. The peaks in the $P_{nl}$ plot remain discernible, effectively capturing the crystal's periodicity and symmetry despite perturbations. In contrast, the $G(d)$ plots become significantly distorted with notable peak shifts. While it is possible to use the autocorrelation techniques to improve $G(d)$ robustness for the similarity calculation~\cite{de2001generalized}, sensitivity to noise remains an issue. Therefore, it is safe to conclude that  $P_{nl}$ provides superior robustness to noise and structural variations compared to $G(d)$. 

In modern materials discovery, generative models~\cite{DiffCSP, levy2024symmcd, zeni2025generative} are frequently employed to rapidly generate candidate crystal structures. Subsequently, duplicate structures need to be removed using real-space structure matching algorithms, such as the StructureMatcher module in \texttt{pymatgen}~\cite{pymatgen-2013}. This conventional approach involves searching for supercell matches between trial structures and verifying atomic coordinate equivalence within specified tolerance thresholds. However, this process can be computationally inefficient for structures with large unit cells and becomes cumbersome due to the sensitivity of results to tolerance parameter choices. In contrast, direct comparison of power spectrum $P_{nl}$ offers a more straightforward and robust alternative for structure matching tasks. A systematic evaluation of the utility of $P_{nl}$ for crystal structure matching will be pursued in future work.

\subsection{Crystal Reconstructions}
Next, we test whether one can reconstruct the crystal when a $P_{nl}$ pattern is known. This is analogical to the single crystal x-ray structure determination problem, where the goal is to recover the atomic positions and cell parameters from the measured diffraction intensities. To achieve this, we can use an optimization framework to iteratively adjust the crystal structure until its power spectrum $P_{nl}$ matches the target pattern with the following minimization objective function,

\begin{figure*}[htbp]
    \centering
    \includegraphics[width=0.98\textwidth]{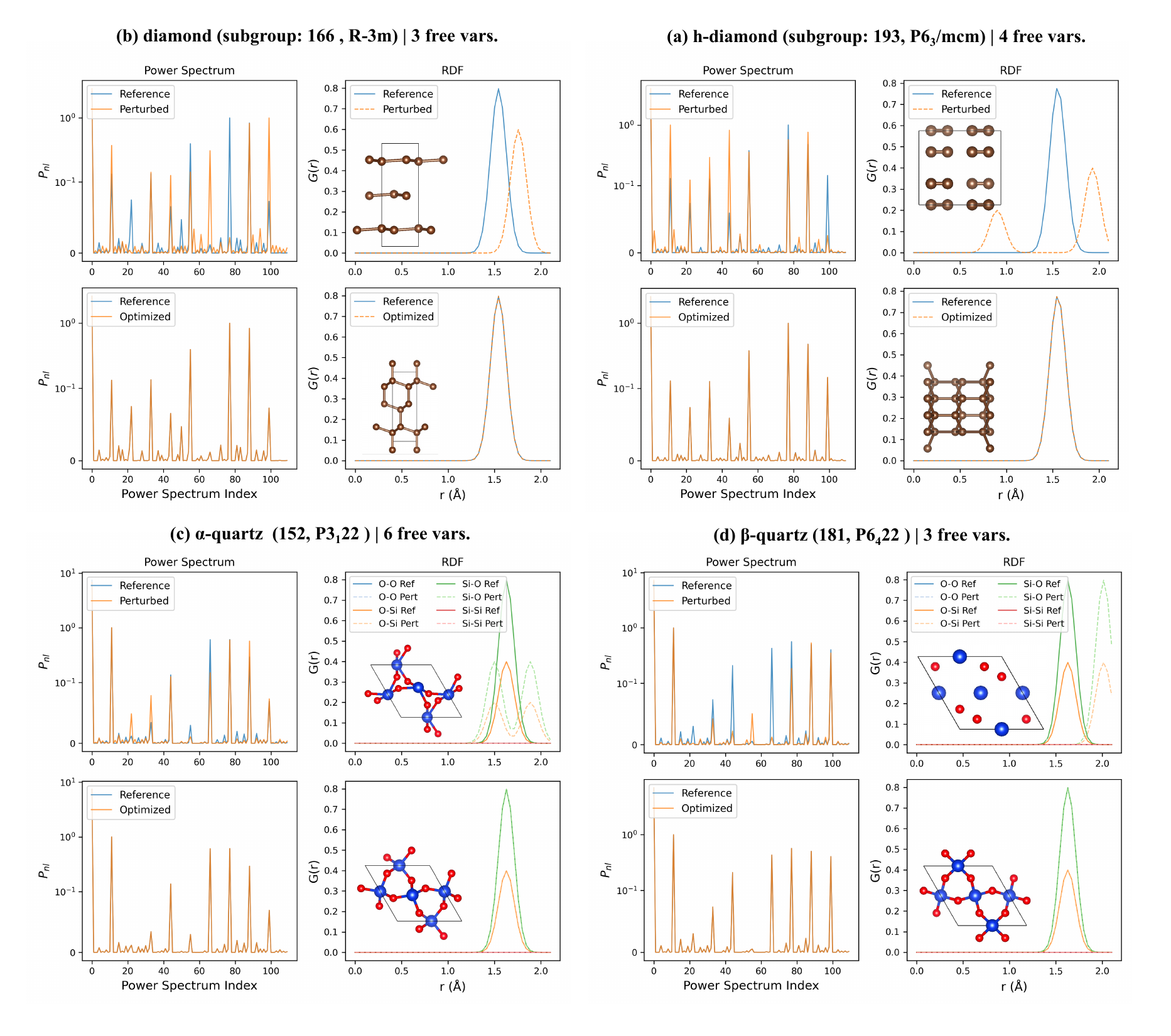}
    \vspace{-3mm}
        \caption{Crystal reconstruction for representative structures including (a) diamond, (b) h-diamond, (c) $\alpha$-quartz, and (d) $\beta$-quartz. }
    \label{fig:recon}
\vspace{-5mm}
\end{figure*}

\[
F_\text{obj.} = \sum_{nl}\|P_{nl}(\mathbf{X})- P^{\text{ref.}}_{nl}\|^2, 
\]

where $\mathbf{X}$ represents the atomic positions and cell parameters of the structure candidate.

However, there is a possibility of having multiple solutions that yield the same $P_{nl}$ values due to the non-uniqueness of the inverse problem \cite{kazhdan2003rotation}. To enhance the robustness of the reconstruction, we include the radial distribution function $G(r)$, which is defined as the sum of the intensities over all directions at each binned distance $r$. Similar to $P_{nl}$, we consider $G(r)$  within a short distance cutoff because it is assumed to provide a smooth structural description that is not sensitive to local atomic changes.

In this case, the optimization objective function can be defined as a mean squared error (MSE) loss function that combines both the power spectrum and RDF representations:

\begin{equation}\label{eq:opt}
F_\text{obj.} = \sum_{nl}\|P_{nl}(\mathbf{X})- P^{\text{ref.}}_{nl}\|^2 + \sum_k ||G(r_k, \mathbf{X}) - G^\text{ref.}(r_k)||^2,
\end{equation}

where $P^{\text{ref.}}_{nl}$ is the target power spectrum pattern, and $G^\text{ref.}(r)$ is the target RDF pattern. 

This can still be a very challenging optimization problem when the dimension of $\mathbf{X}$ is large. To ease the problem, we leverage the symmetry information and optimize the crystal structure in a reduced space assuming that the space group and Wyckoff choices are known. During optimization, one treats the space group and Wyckoff choices as the conditioned discrete variables and then optimizes the reduced variables (e.g. free cell parameters and xyz coordinates) in the continuous space. The goal of the optimization procedure is to adjust reduced variables to minimize this MSE loss function value, thereby steering structures toward desired crystal structures.

To test the feasibility of using optimization for the reconstruction task, we consider four representative prototypes---diamond, h-diamond, $\alpha$-quartz, and $\beta$-quartz. For each structure, we compute the target power spectrum $P_{nl}^{\mathrm{ref.}}$ from the reference crystal. An initial structure is generated by applying random perturbations to the reduced variables (including both cell parameters and xyz coordinates). Specifically, the cell parameters are mutated by a relative degree $d_\mathrm{lat}$ by scaling the independent lattice parameters with a random factor $1+ud_\mathrm{lat}$ with $u \in [-0.5, 0.5)$ while enforcing the constraints of the lattice type (e.g., cubic $a=b=c$, hexagonal $a=b$ and $\gamma=120^\circ$). The coordinate perturbation is controlled by $d_{\mathrm{coor}}$, which denotes a fixed Cartesian displacement magnitude (in \AA): each Wyckoff generator is shifted by a random vector with $\|\Delta \mathbf{r}\|=d_{\mathrm{coor}}$. In the standard setting, diamond has only 1 free variable on the cell parameters, and h-diamond has 3 free variables (2 on the cell parameters, and 1 on the Wyckoff position). We purposely choose their subgroup settings ($P6_3/mcm$ for h-diamond and $R$-3$m$ for diamond) to allow more variables to increase the optimization challenge. 
After perturbation, we then minimize the objective function according to Eq.~(\ref{eq:opt}), using the standard optimization algorithms of Nelder--Mead and L-BFGS in SciPy~\cite{scipy}.

Fig.~\ref{fig:recon} demonstrates the results before and after optimization for all cases with the following choices: $(0.5, 0.5)$ for diamond, $(0.2, 1.0)$ for h-diamond, $(d_\mathrm{lat}, d_\mathrm{coor})=(0.1, 0.9)$ for $\alpha$-quartz, $(0.2,1.8)$ for $\beta$-quartz. In all four cases, we display the calculated $P_{nl}$ and $G(r)$ values for both initial (upper plot) and optimized structures (lower plot). Clearly, both $P_{nl}$ and $G(r)$ are found to match the target patterns after optimization, despite that their initial patterns are notably different. These encouraging results demonstrate that, with a reasonable initial guess of the space group and Wyckoff xyz, it is feasible to reconstruct the underlying crystal structure from its reciprocal space $P_{nl}$ and real space RDF $G(r)$.

However, it is important to note that the optimization landscape is intrinsically nonconvex. For stronger perturbations with larger $(d_\mathrm{lat}, d_\mathrm{coor})$ values, we found the optimization is more likely to become trapped in local minima. When more reduced variables are involved (e.g., the case of $\alpha$-quartz has 6 variables in Fig.~\ref{fig:recon}c), the simple optimization is successful only when the initial guess is very close to the ground truth structures. In such cases, global optimization or multi-start strategies (e.g., basin hopping~\cite{Wales-JPCA-1997}) may improve robustness when the initial structure is far from the desired configuration. 

Recently developed generative models for materials discovery iteratively refine random noise into realistic structures by sampling from a learned distribution \cite{DiffCSP, levy2024symmcd, zeni2025generative}. While such models separately treat atomic coordinates and cell parameters to account for rotational and translational invariance, they may overfit to training data without explicitly enforcing physical constraints. The proposed $P_{nl} + G(r)$ representation can serve as an additional constraint to guide the generative process, ensuring the physical validity of synthesized crystals. 
Future work will focus on integrating this representation into generative models and exploring its applications in crystal property prediction and inverse design.

\section{Conclusions}\label{sec5}
In this work, we present a representation of crystal structures using 4D reciprocal space that naturally captures their periodicity and symmetry. We further introduce a power spectrum representation, constructed from spherical harmonics and radial basis functions, to achieve rotational invariance. This representation can be computed efficiently while maintaining rotational and translational invariance. Our numerical experiments demonstrate that the power spectrum representation provides a robust and smooth characterization of structures even under significant perturbations. When combined with the real space radial distribution function, it enables crystal structure reconstruction through optimization techniques. These results suggest that power spectrum representation of 4D reciprocal space may be used in the modern materials discovery workflow such as crystal matching and conditioned material generation tasks.

\section*{Acknowledgments}
This research was sponsored by the U.S. Department of Energy, Office of Science, Office of Basic Energy Sciences, and the Established Program to Stimulate Competitive Research (EPSCoR) under the DOE Early Career Award No. DE-SC0024866, and the UNC Charlotte's seed grant from data science.

\section*{Data availability}
The source codes used in this study are available in \url{https://github.com/MaterSim/ReciprocalXtal}.

\section*{Conflict of interest}
All authors declare that they have no conflict of interest.

\nolinenumbers
\section*{REFERENCES}
\bibliography{ref}

\end{document}